# Group Methodologies and Simulations for the Development of Transversal Skills: A Pilot Study on Health Sciences Higher Education

**Laura San Martín Galindo** [ab]**, Juan José Cabrera-Martinez** [b]**, Camilo Abalos-Labruzzi** [b] **and José Gómez-Galán** [c]

**Abstract**: One of the methodologies based on group dynamics is Role Playing (RP). This method consists on the simulation of a real situation, allowing its study and understanding. Knowledge and technical skills are not the only prerequisites for proper practice in health sciences. RP has been used as Communication Skills Training (CST) amongst health professionals. In the teaching of odontology and stomatology, object of our research, dental assistance brings up situations where the professional must develop transversal skills, which improve the interaction with the patient and the dental treatment itself. The aim of this study is to evaluate the efficacy of the proposed RP methodology through the students' opinion. The study sample is made out of dental students (n=80), all of them on the 4th year at the College of Dentistry (University of Seville, Spain). The students who took part in the activity considered the incorporation of RP in the syllabus as relevant, though further study should be considered in order to analyze the efficacy of this teaching methodology in depth.

## 1. Introduction

The relevance of teaching methods for the quality of learning has been widely demonstrated in multiple studies (Ramsden, 2003; Addison, Burgess,

---

[a] University of Harvard (United States); [b] Universidad de Sevilla (Spain); [c] Universidad Metropolitana (Puerto Rico, United States). Correspondence: Laura San Martin, Department of Oral Health Policy and Epidemiology, Harvard School of Dental Medicine, 188 Longwood Avenue, Boston, MA 0215 (United States). laura_sanmartin@hms.harvard.edu.

Steers & Trowell, 2010; Cobb, & Jackson, 2012; Davies, Jindal-Snape, Collier, Digby, Hay & Howe, 2013). In Higher Education the traditional position has been the use of exhibition techniques, also known as expository lessons and mass instruction ((Ellington & Race, 1993), which led to the discarding of other methodological alternatives and strategies, that while being suitable for other educational levels, were considered unsuitable for the university. Nevertheless, the technological revolution brought about by ICTs has created the need for the transformation of traditional processes of university education to adjust to the new social reality in which students are immersed (Gómez-Galán & Mateos, 2002; Gómez-Galán, 2002; 2014a).

Certainly, the training of future professionals in the various fields of knowledge that are prepared at University cannot only be limited to the theory or practice proper to their specialty, but it is also necessary to acquire different skills and abilities for their work in the field of society. Thus, the employment by university professors of other methodologies parallel to exhibition techniques can offer unquestionable advantages. The application of group dynamics, socialization techniques, discovery learning, autonomous work and self-learning strategies, alternative methodologies and the use of ICTs for the development of conceptual, procedural and attitudinal contents, etc., will certainly help a university student to take responsibility, make decisions, enhance the processes of study and research, etc. and, above all, motivate and enhance their creativity (Gómez-Galán, 2014b; Cropley, 2015).

The use of multiple teaching methods is also positive for the teacher, because it forces them to assess the intervention processes, analyze the educational possibilities of the means and resources at their disposal and study the processes of teaching and learning in relation to pursued objectives and characteristics of contents to offer. Also, group dynamics and simulations gain special importance among these methodologies since they are critical to the quality of learning in the academic sector (Jones, 1984; Thorley, Gregory, & Gregory, 1994; Recker, Govindaraj, & Vasandani, 1998; Gómez-Galán, 2002; Hertel & Millis, 2002; Lean, Moizer, Towler, & Abbey, 2006; Hoffman, Wilkinson, Xu, & Wiecha, 2014).

One of the methodologies based on group dynamics is role playing (RP), which in the academic field is defined as a teaching method based on group dynamics, which uses a simulation focused on the interaction between students with different roles in several circumstances, generating meaningful learning close to real life (Van Ments, 1989; DeNeve & Heppner, 1997; Burns & Gentry, 1998; Martínez-Riera, J. R., Sanjuán, A., Cibanal, L., & Pérez-Mora, 2011). RP works the teaching–learning process acquiring the skills through proposed simulated situations. The students face with unexpected events and seek out the best solution. The Retrospective thinking through a discussion has particular relevance for feedback, especially devised when trying to develop the communicative skills. This is a constructivist-learning model, opposed to passively receiving information (Gil & Guzmán, 2005).

The design of simulated situations should be selected according to whether the educational goal addresses knowledge, attitudes or skills. In the acquisition of communication skills, repeated opportunities with feedback permit the students to achieve an effective communication. The objective of the simulation is to assess the student's ability to provide information, involve them in making professional decisions and eliminates barriers in communication (Jones, 1982; Mayer, 1993; Maier, 2002). A variety of methods are used to accomplish communication objectives. RP has been used as Communication Skills Training (CST) amongst health professionals. Knowledge and technical skills are not the only prerequisites for proper practice in health sciences, and effective communication is an element related with good practices and its perception from patients, having multiple impacts on various aspects of health outcomes (Spitzberg, 2013; San Martin, Utrilla & Mediavilla, 2014).

The field of medicine has generated a number of research studies in communication skills training. Teaching recommendations for effective teaching have been identified as a skills-based approach, self-assessment by students, small groups for optimal learning and multidisciplinary teaching staff involved (1998; Maguire, Fairbairn & Fletcher, 1998; Silverman, Kurtz & Draper, 2013; Naghavi, Anbari, Saki, Mohtashami, Lashkarara, & Derik, 2014). The Association of American Medical Colleges, the Accreditation Council for Graduate Medical Education, and others have suggested for medical educators to define, teach, and evaluate communication skills for physicians in training (Haq, Steele, Marchand, Seibert & Brody, 2004). Mixed methodologies have been used successfully on the nursery field, with combined skills aiming towards the improvement of the clinical practice and the learning of an effective communication with patients (Mullan & Kothe, 2010).

In the field of dentistry, the importance of behavioral sciences, and in particular communication skills, was formally recognized by the General Dental Council's publishing guidelines for its inclusion in teaching in dental schools (Hananah, 2004). Yoshida, Milgrom and Coldwell (2002), in a review conducted in the U.S. and Canadian dental schools, found that only one-third of them had courses focusing specifically on interpersonal communication. The authors found that the programs took the form of passive learning rather than active skills-based practice using simulated, concluding that there is a current lack of teaching communication skills in North America. On the other hand in Europe, behavioral sciences teaching, including communication skills, show considerable variation with regard to the contents, duration, the credentials of instructors involved and the teaching methods employed (Goldrick, 1999).

There are not too many published studies about the evaluation of RP as an educative method compared to other teaching methodologies and little consensus across the professions regarding how it should be assessed and embedded throughout the curriculum (Hargie, Dickson, Boohan, & Hughes, 1997; Hannah, Millichamp, & Ayers, 2004; Hargie, Boohan, McCoy, & Murphy, 2010).

The aim of our study, this way, is the evaluation and the inclusion of a practical teaching methodology into the current teaching of the Dentistry Degree in Spain, focused on the simulation and interaction with patients and the teamwork. This pilot study focuses on the development of the communication skills, making a difference with previous studies focused on the acquisition of technical skills in clinical practice (El Tantawi, Abdelaziz, AbdelRaheem, & Mahrous, 2014). We propose this pilot study as part of our research strategy which will serve as a basis for future study questions, besides allowing the comparison of results. We seek to deepen our understanding of new teaching methods in health sciences higher education.

## 2. Method

### 2.1. Materials and Participants

Many audiovisual devices as slides presentations and recordings have been used to make the theoretical introduction of the activity in order to provide an easier understanding and development.

The study sample is made out of 80 students, 62 female and 20 male, with ages from 21 to 25 years old, all of them dental students on the 4th year at College of Dentistry, University of Seville (Spain) (See figure 1).

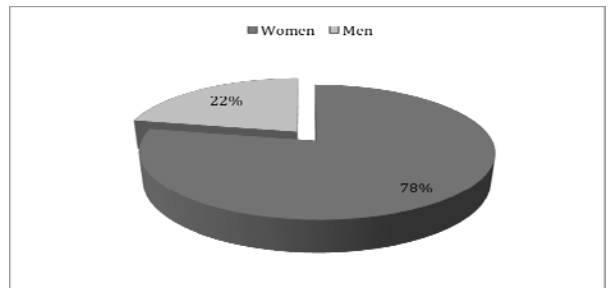

*Figure 1. Percentage of gender distribution on the sample.*

### 2.2. Design

This is a cross-sectional descriptive study. A research design with quantitative methodology has been used on the development of this work. The evaluation of the teaching methodology has been conducted through a questionnaire with 10 items, 5 items for pre-test and 5 items for post-test under assertions or opinions by using a Likert scale.

Afterwards, a descriptive study of the answers has been designed, building distributions of frequencies and an analysis of relations with the variables with an application of contingency tables through the Chi–square test. For this analysis, the program *Statistical Package for the Social Sciences version 20 (SPSS®)* has been used.

### 2.3. Procedure

Phase 1: Preparatory phase. *Training and formation of students in the classroom, group shaping and role assignment.*

This phase includes a preliminary needs assessment of the group, a brief overview of the cognitive approach to the content and the creation of an effective learning environment. The participating instructors came with the updated content about the topic their activity was going to be about and they exposed the general rules to start with the activity. In groups, students had 10 minutes to organize the different roles and prepare the play, with those roles described by the participants. All students should be involved and ground rules should be set.

Phase 2: Dramatization phase. *Group Simulations, assessment and co-evaluation (evaluation between peers) of the group simulation.*

The students made a short and improvised staging, of around 15 minutes, which was recorded for a subsequent analysis and assessment. The observers of the staging took note on an observation scale of the main elements that a dentist should value when interacting with patients and the level of development of the communication skills and oral expression skills were valued. The role for the instructor during this phase is to monitor the simulation, deciding when to intervene.

Phase 3: Debate Phase. *Reflexive / analytic debate about the staging and subsequent evaluation.*

When the staging is finished, the instructor started a deep and reflexive debate of around 30 minutes about the staging performed, concluding with feedback and debriefing, in order to understand widely the issue dealt with. This is made in order to create better strategies to face this issue.

### 2.4. Evaluation

In order to achieve a satisfactory analysis of the activity, the academic performance and the monitoring of the project have been evaluated.

*Evaluation of the academic performance*: The students´ participation counts as positive for the final marks, adding one point to the practical evaluation of the given subject.

*Evaluation of the monitoring of the project*: When the activity is finished, the students have fulfilled the evaluation questionnaire made out of items concerning the interest, organization, content, achieved learning, motivation capacity on the activity and application of the activity on the working field.

### 3. Results

*3.1. Pre-test*

80% of students polled considered as relevant the acquisition of skills and professional attitudes of behavior (m=4.22) and 95% agreed with the usefulness and relevance for a professional career (m=4.72).

15% of the students polled argued that acquired knowledge over the degree was enough (m=2.83) and 55% remarked that they had improved their interpersonal skills over the degree years (m=3.55). (See table 1)

| Item | | f | (%) | mean |
|---|---|---|---|---|
| **1. I think it is relevant to acquire skills and professional attitudes of behavior which allow an effective and adequate both with patients and the team.** | | | | |
| 1 | Strongly disagree | 0 | 0 | |
| 2 | Disagree | 0 | 0 | |
| 3 | Neither agree nor disagree | 16 | 20 | |
| 4 | Agree | 24 | 30 | |
| 5 | Strongly agree | 40 | 50 | |
| **Total** | | 80 | 100 | 4.22 |
| Item | | f | (%) | mean |
| **2. I think the training and formation received over the dentistry degree are enough to acquire interpersonal skills.** | | | | |
| 1 | Strongly disagree | 12 | 15 | |
| 2 | Disagree | 20 | 25 | |
| 3 | Neither agree nor disagree | 36 | 45 | |
| 4 | Agree | 8 | 10 | |
| 5 | Strongly agree | 4 | 5 | |
| **Total** | | 80 | 100 | 2.83 |
| **3. I have acquired interpersonal skills over the previous academic years.** | | | | |
| 1 | Strongly disagree | 16 | 20 | |
| 2 | Disagree | 4 | 5 | |
| 3 | Neither agree nor disagree | 24 | 30 | |
| 4 | Agree | 32 | 40 | |
| 5 | Strongly agree | 4 | 5 | |
| **Total** | | 80 | 100 | 3.28 |
| **4. My interpersonal skills have improved over the degree years.** | | | | |
| 1 | Strongly disagree | 0 | 0 | |
| **2** | Disagree | 8 | 10 | |
| **3** | Neither agree nor disagree | 28 | 35 | |
| **4** | Agree | 40 | 50 | |
| **5** | Strongly agree | 4 | 5 | |
| Total | | **80** | **100** | **3.55** |
| **5. I think that the knowledge about the patient treatment and teamwork is useful and relevant for a professional career.** | | | | |
| **1** | Strongly disagree | 0 | 0 | |
| **2** | Disagree | 0 | 0 | |
| **3** | Neither agree nor disagree | 4 | 5 | |
| **4** | Agree | 12 | 15 | |
| **5** | Strongly agree | 64 | 80 | |
| **Total** | | **80** | **100** | **4.72** |

*Table 1. Frequency, percentage and items average (pre-test).*

### 3.2. Post-test

100% of the students polled agree (60%) or very much agreed (40%) with having found the goal of RP in dentistry. The punctuation obtained about the clarity (m=4.50), organization (m=4.35) and adaptation to case studies (m=4.55) have shown a superior average to 4 on all items. The link of the activity with the professional career of dentist has shown the highest value of all punctuations with an average of 4.60 (see table 2)

| Item | | f | (%) | media |
|---|---|---|---|---|
| **6. The goal of role playing during the training of the dentistry degree has been identified.** | | | | |
| **1** | Strongly disagree | 0 | 0 | |
| **2** | Disagree | 0 | 0 | |
| **3** | Neither agree nor disagree | 0 | 0 | |
| **4** | Agree | 48 | 60 | |
| **5** | Strongly agree | 32 | 40 | |
| **Total** | | **80** | **10** | **4.40** |
| **7. The key issues of the topic have been clearly presented.** | | | | |
| **1** | Strongly disagree | 0 | 0 | |
| **2** | Disagree | 0 | 0 | |
| **3** | Neither agree nor disagree | 0 | 0 | |
| **4** | Agree | 40 | 50 | |
| **5** | Strongly agree | 40 | 50 | |
| **Total** | | **80** | **10** | **4.50** |
| **8. The organization of the contents has been noticed.** | | | | |
| **1** | Strongly disagree | 0 | 0 | |
| **2** | Disagree | 0 | 0 | |
| **3** | Neither agree nor disagree | 12 | 15 | |
| 4 | Agree | 28 | 35 | |
| 5 | Strongly agree | 40 | 50 | |
| **Total** | | 80 | 10 | **4.35** |
| **9. The case studies have been appropriate.** | | | | |
| 1 | Strongly disagree | 0 | 0 | |
| 2 | Disagree | 0 | 0 | |
| 3 | Neither agree nor disagree | 0 | 0 | |
| 4 | Agree | 36 | 45 | |
| 5 | Strongly agree | 44 | 55 | |
| Total | | 80 | 10 | **4.55** |
| **10. The link with the professional career of a dentist has been clearly stated.** | | | | |
| 1 | Strongly disagree | 0 | 0 | |
| 2 | Disagree | 0 | 0 | |
| 3 | Neither agree nor disagree | 0 | 0 | |
| 4 | Agree | 32 | 40 | |
| 5 | Strongly agree | 48 | 60 | |
| **Total** | | 80 | 10 | **4.60** |

*Table 2. Frequency, percentage and items average (post-test)*

In the test made through contingency tables with the chi-square sample, the result is that the proportion of both sexes has been obtained in a generic way (women 78%, men 22%) and it is seen in the answers of these individuals when agreeing or disagreeing on the matters (see tables 3 and 4, figures 2-11)

| Gender | | Item 1 | Item 2 | Item 3 | Item 4 | Item 5 |
|---|---|---|---|---|---|---|
| **Male** | **Mean** | 4.39 | 2.50 | 2.72 | 3.72 | 4,56 |
| | **N** | 18 | 18 | 18 | 18 | 18 |
| | **SD** | 0.60 | 1.09 | 1.27 | 0.75 | 0,71 |
| **Female** | **Mean** | 4.27 | 2.69 | 3.15 | 3.44 | 4,81 |
| | **N** | 62 | 62 | 62 | 62 | 62 |
| | **SD** | 0.83 | 1.00 | 1.18 | 0.74 | 0,47 |
| **Total** | **Mean** | 4,30 | 2,65 | 3,05 | 4,75 | 4,40 |
| | **N** | 80 | 80 | 80 | 80 | 80 |
| | **SD** | 0,79 | 1,02 | 1,21 | 0,75 | 0,54 |

*Table 3. Summary of dispersion measures according to gender (pre-test)*

| Gender | | Item 6 | Item 7 | Item 8 | Item 9 | Item 10 |
|---|---|---|---|---|---|---|
| **Male** | **Mean** | 4,56 | 4,33 | 4,33 | 4,33 | 4,72 |
| | **N** | 18 | 18 | 18 | 18 | 18 |
| | **SD** | 0,51 | 0,49 | 0,59 | 0,49 | 0,46 |
| **Female** | **Mean** | 4,35 | 4,55 | 4,35 | 4,55 | 4,56 |
| | **N** | 62 | 62 | 62 | 62 | 62 |
| | **SD** | 0,48 | 0,50 | 0,77 | 0,50 | 0,50 |

|       | Mean | 4,40 | 4,50 | 4,35 | 4,55 | 4,60 |
|-------|------|------|------|------|------|------|
| Total | N    | 80   | 80   | 80   | 80   | 80   |
|       | SD   | 0,49 | 0,50 | 0,73 | 0,50 | 0,49 |

*Table 4. Summary of dispersion measures according to gender (Post-test)*

Figures 2-6. Scores of contingency tables, chi-square test for items on pre-test:

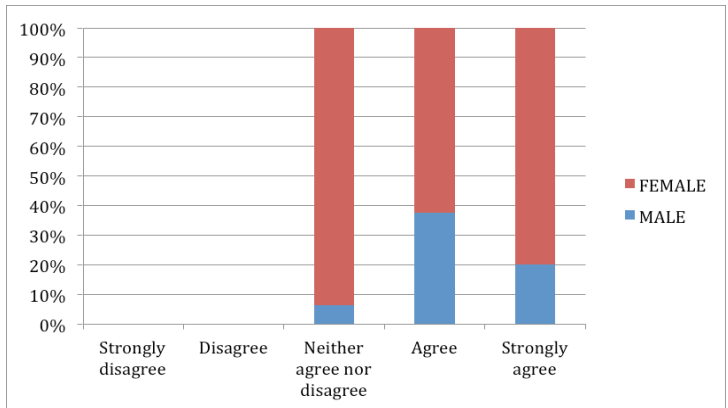

*Figure 2. Scores of contingency tables, chi-square test for items on item 1.*

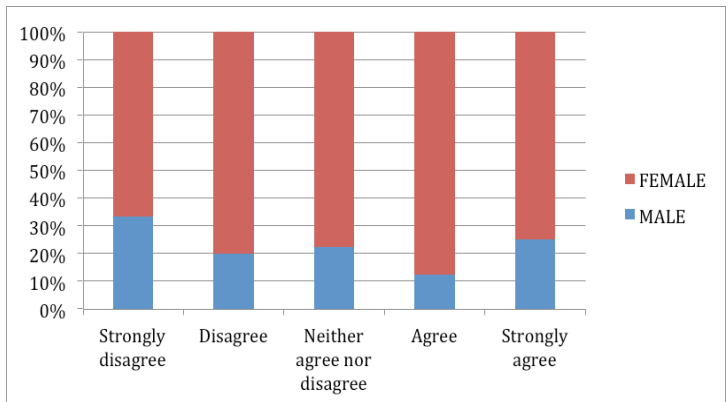

*Figure 3. Scores of contingency tables, chi-square test for items on item 2.*

101

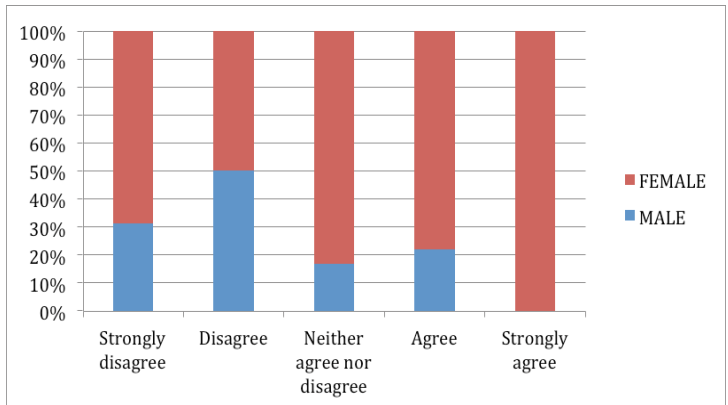

*Figure 4. Scores of contingency tables, chi-square test for items on item 3.*

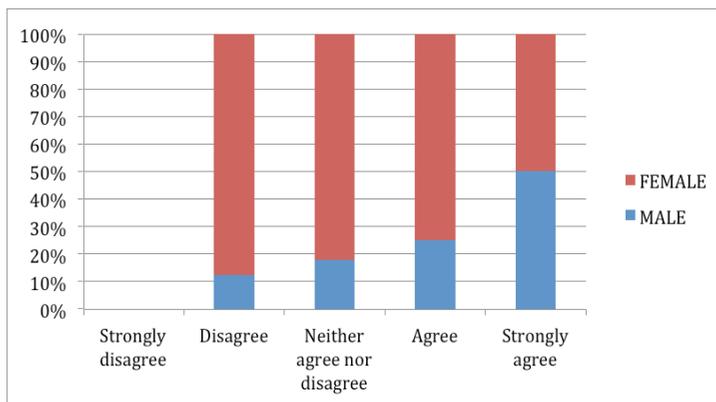

*Figure 5. Scores of contingency tables, chi-square test for items on item 4.*

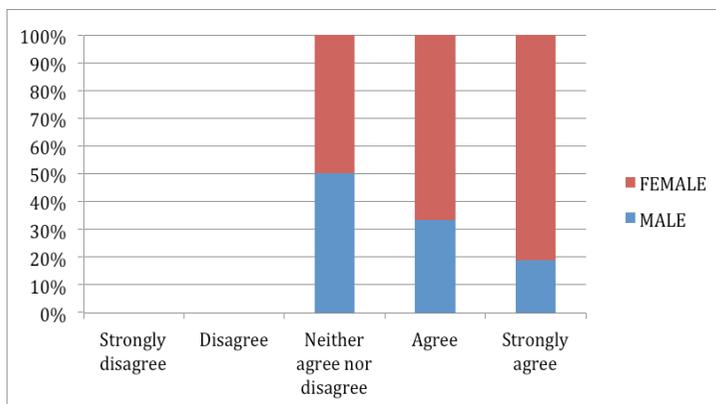

*Figure 6. Scores of contingency tables, chi-square test for items on item 5.*

Figures 7-11. Scores of contingency tables, chi-square test for items on post-test:

102

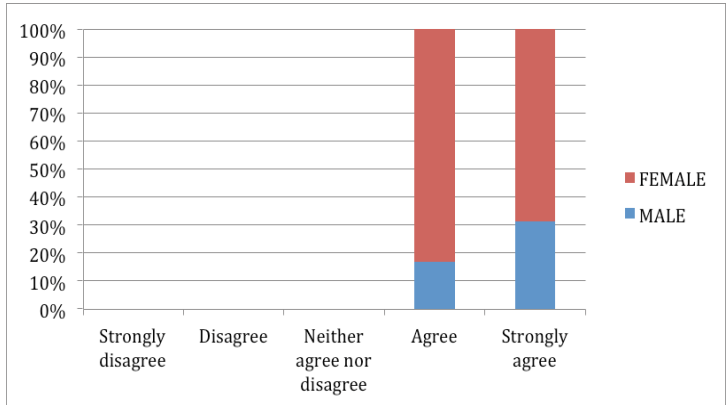

*Figure 7. Scores of contingency tables, chi-square test for item on item 6.*

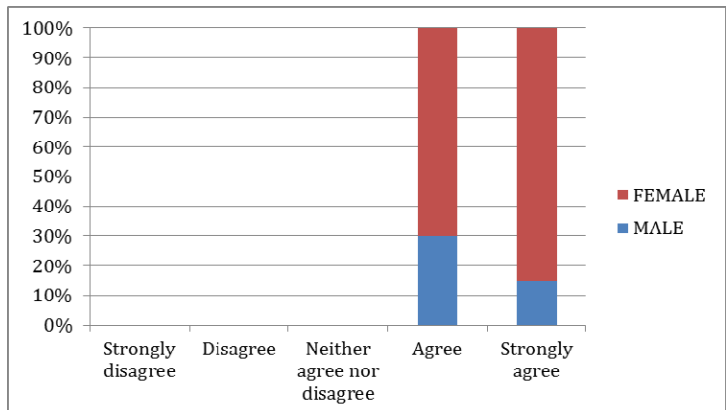

*Figure 8. Scores of contingency tables, chi-square test for items on item 7.*

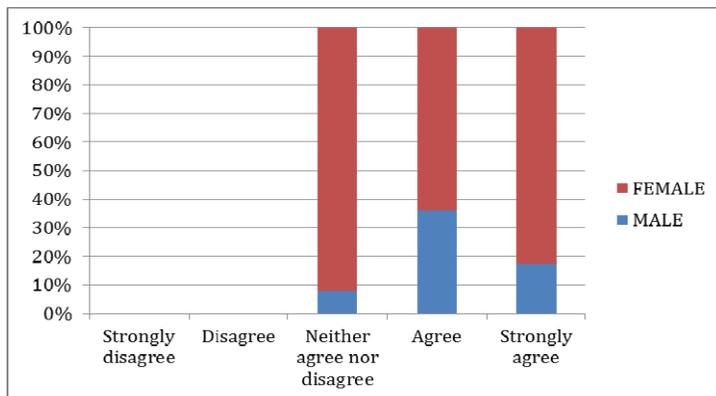

*Figure 9. Scores of contingency tables, chi-square test for items on item 8.*

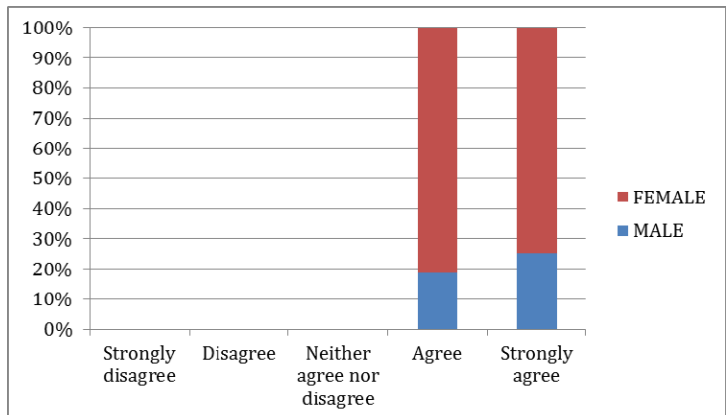

*Figure 10. Scores of contingency tables, chi-square test for items on item 10.*

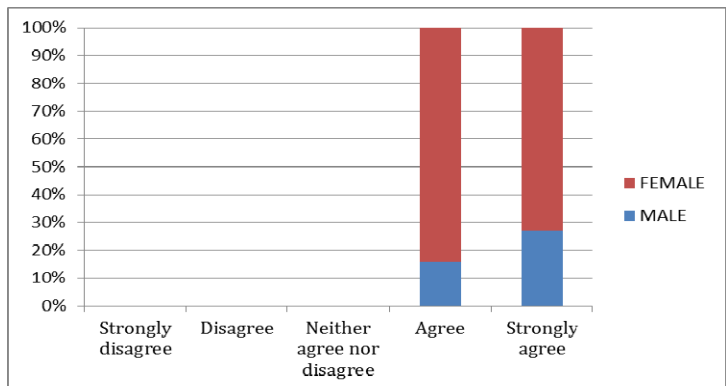

*Figure 11. Scores of contingency tables, chi-square test for items on item 11.*

## 4. Discussion and Conclusions

The participation percentage in this research was 100%, as it was designed as an activity inside the program of the subject and it was taken into account in the final evaluation. Some authors argue that RP may cause some rejection on students. Stevenson and Sander (2002) informed that this is the less accepted teaching method in 32% of medical students, remarking the ineffectiveness of this methodology or personal reasons like embarrassment when staging in front of their peers.

The results obtained on this study (95%) differ from these results but they are similar to Knowles, Holton and Swanson (2011), in which 71% of the participants considered RP as a useful and effective technique for the learning of such skills like communication. The School of Dentistry of Birmingham (United Kingdom) reported a positive assessment (69,7%) on the role-play teaching methodology showing its relevance for a professional career (Croft, White, Wiskin, & Allan, 2005).

The gradual introduction of the activity is very important, especially if the group of students is not familiar with the experimental way of working. Introducing RP may involve the resistance or anxiety from some students (Burns & Gentry; 1998; Van Ments, 1989). RP situations, dissociated from a clinical case and introduced previously, help the students to lose their fear and reticence to this method, increasing their participation on subsequent approaches, where a high participation to acquire a particular skill is greatly advisable (Christiaens & Baldwin, 2002).

Despite the results obtained in previous studies and in this pilot study which shows positive attitudes toward RP, there is a need to make more studies with an appropriate design which evaluate the acquisition of skills after the use of simulated patients and RP in a series of different situations (Lane & Rollnick, 2007). RP experiences in the dental curriculum tended to occur on a one-course only basis, so there is no opportunity for students to learn gradually and to increase the complexity of the activity (Yoshida, Milgrom, & Coldwell, 2002). The results of this pilot study were only gathered from one year, so a comparative study with students' prior experiences is not possible. The keys to successful teaching include assessment of the learner and the educator's ability to provide constructive and goal-directed feedback (Jackson & Back, 2011).

In this study, the opinion of students about the usefulness of RP has been analyzed based on a Likert scale. Other authors have used scales (Communication Skills Attitude Scale CSAS), which offer a standardized evaluation of the level of acquisition of skills in the teaching-learning process. This way, Laurence, Bertera, Feimster, Hollander and Stroman (2012) made an adaptation of the SAC especially designed for students of dentistry (Dental Communication Skills Attitude Scale DCSAS), concluding that it is a useful tool for the evaluation of attitudes towards the learning of communication skills among dental students.

Besides, RP is widely used as an educational method for Communication Skills Training (CST), it should take a further look at the analysis of factors, which may influence the activity, and at the evaluation of the methodology for the development of specific skills in the health area (Kruijver, 2000; Nestel, 2007; Okuda, 2009).

Concluding, the inclusion of group methodologies has enabled us to improve the learning process and the acquisition of transversal skills for students, developing communication skills, teamwork and decision-making skills, which may have an effect on an improvement on the patient care. New research should be carried out about the RP as Communication Skills Training (CST) to dentistry, in order to deeply analyze the effectiveness of the teaching method.

It must be borne in mind, first, that innovative proposals in the field of teaching methodology in Higher Education are not only desirable but completely necessary in today's society, where patterns of transmission of information and communication processes have changed profoundly. The application of new pedagogical paradigms such as formative assessment, interaction-based teacher-student communication, teaching methodologies that enhance participation and activity of students, integration of the educational use of ICTs, etc., should represent the basis of new Higher Education structures (Gómez-Galán & Mateos, 2002). We are heading towards the development of more open and democratic pedagogies that encourage responsibility and students´ self-learning. This transformation, based on teacher training, strengthens the figure and the professionalism of the teaching function. In the case of Health Sciences the training of future professionals will improve in this field and, consequently, the care provided to their patients.

**References**

Addison, N. Burgess, L. Steers, J., & Trowell, J. (2010). *Understanding Art Education: Engaging Reflexively with Practice*. London: Routledge.

Burns A. C., & Gentry J. W. (1998). Motivating Students to Engage in Experiential Learning: A Tension-to-Learn Theory. *Simulation and Gaming*, 29, 133-15.

Christiaens, G., & Baldwin, J. H. (2002). Use of Didactic Role-Playing to Increase Student Participation. *Nurse Educator*, 27(6), 251-254.

Cobb, P., & Jackson, K. (2012). Analyzing Educational Policies: A Learning Design Perspective. *Journal of the Learning Sciences*, 21(4), 487-521.

Croft, P., White, D. A., Wiskin, C. M. D., & Allan, T. F. (2005). Evaluation by Dental Students of a Communication Skills Course Using Professional Role-Players in a UK School of Dentistry. *European Journal of Dental Education*, 9(1), 2-9.

Cropley, D. H. (2015). Promoting Creativity and Innovation in Engineering Education. *Psychology of Aesthetics, Creativity, and the Arts*, 9(2), 161.

Davies, D., Jindal-Snape, D., Collier, C., Digby, R., Hay, P., & Howe, A. (2013). *Creative Learning Environments in Education: A Systematic Literature Review*. Thinking Skills and Creativity, 8, 80-91.

De Neve, K. M., & Heppner, M. J. (1997). Role Play Simulations: The Assessment of an Active Learning Technique and Comparisons with Traditional Lectures. *Innovative Higher Education*, 21(3), 231-246.

Ellington, H. & Race, P. (1993). *Producing Teaching Materials*. London: Kogan Page.

El Tantawi, M. M., Abdelaziz, H., AbdelRaheem, A. S., & Mahrous, A. A. (2014). Using Peer-Assisted Learning and Role-Playing to Teach Generic Skills to Dental Students: The Health Care Simulation Model. *Journal of Dental Education*, 78(1), 85-97.

Gil, D. y Guzmán, M. (2001). *Enseñanza de las Ciencias y la Matemática: Tendencias e Innovaciones*. Madrid: OEI

Goldrick, P., & Pine, C. (1999). Behavioural Sciences: A Review of Teaching of Behavioural Sciences in the United Kingdom Dental Undergraduate Curriculum. *British Dental Journal*, 186(11), 576-580.

Gómez-Galán, J. (2002). Education and Virtual Reality. In N. Mastorakis (ed.). *Advances in Systems Engineering, Signal Processing and Communications*. (pp. 345-350). New York: WSEAS Press.

Gómez-Galán, J. (2014). El Fenómeno MOOC y la Universalidad de la Cultura: Las Nuevas Fronteras de la Educación Superior. *Revista de Curriculum y Formación del Profesorado*, 18(1), 73-91.

Gómez-Galán, J. (2014). Transformación de la Educación y la Universidad en el Postmodernismo Digital: Nuevos Conceptos Formativos y Científicos. In Durán, F. (coord.). *La Era de las TIC en la Nueva Docencia* (pp. 171-182). Madrid: McGraw-Hill

Gómez-Galán, J., & Mateos, S. (2002). Retos Educativos en la Sociedad de la Información y la Comunicación. *Revista Latinoamericana de Tecnología Educativa*, 1(1), 9-23.

Hannah, A., Millichamp, C. J., & Ayers, K. M. (2004). A Communication Skills Course for Undergraduate Dental Students. *Journal of Dental Education*, 68(9), 970-977.

Haq, C., Steele, D. J., Marchand, L., Seibert, C., & Brody, D. (2004). Integrating the Art and Science of Medical Practice: Innovations in Teaching Medical Communication Skills. *Family* Medicine, 36(1), 543-550.

Hargie, O., Boohan, M., McCoy, M., & Murphy, P. (2010). Current Trends in Communication Skills Training in UK Schools of Medicine. *Medical Teacher*, 32(5), 385-391.

Hargie, O., Dickson, D., Boohan, M. & Hughes, K. (1997). A Survey of Communication Skills Training in UK Schools of Medicine: Present Practices and Prospective Proposals. *Medical Education*, 32(1), 25-34.

Hertel, J. P., & Millis, B. J. (2002). *Using Simulations to Promote Learning in Higher Education*. Sterling, VA: Stylus Publishing.

Hoffman, M., Wilkinson, J. E., Xu, J., & Wiecha, J. (2014). The Perceived Effects of Faculty Presence vs. Absence on Small-Group Learning and Group Dynamics: A Quasi-Experimental Study. *BMC Medical Education*, 14(1), 258.

Jackson, V. A., & Back, A. L. (2011). Teaching Communication Skills Using Role-Play: An Experience-Based Guide for Educators. *Journal of Palliative Medicine*, 14(6), 775-780.

Jones, K. (1982). *Simulations in Language Teaching*. Cambridge: Cambridge University Press.

Jones, K. (1984). Simulations versus Professional Educators. In D. Jaques &

E. Tippen (Eds.), *Learning for the Future with Games and Simulations* (pp. 45–50). Loughborough, UK: SAGSET/Loughborough, University of Technology.

Knowles, M. S., Holton, E. F., & Swanson R. A. (2011). *The Adult Learner: The Definitive Classic in Adult Education and Human Resource Development*. London: Elsevier.

Kruijver, I. P., Kerkstra, A., Francke, A. L., Bensing, J. M., & van de Wiel, H. (2000). Evaluation of Communication Training Programs in Nursing Care: A Review of the Literature. *Patient Education and Counseling*, 39(1), 129-145.

Lane, C., & Rollnick, S. (2007). The Use of Simulated Patients and Role-Play in Communication Skills Training: A Review of the Literature to August 2005. *Patient Education and Counseling*, 67(1), 13-20.

Laurence, B., Bertera, E. M., Feimster, T., Hollander, R., & Stroman, C. (2012). Adaptation of the Communication Skills Attitude Scale (CSAS) to Dental Students. *Journal of Dental Education*, 76(12), 1629-1638.

Lean, J., Moizer, J., Towler, M., & Abbey, C. (2006). Simulations and Games Use and Barriers in Higher Education. *Active Learning in Higher Education*, 7(3), 227-242.

Maguire, P., Fairbairn, S., & Fletcher, C. (1986). Consultation Skills of Young Doctors: Benefits of Feedback Training in Interviewing as Students Persist. *British Medical Journal*, 292, 1573-1578.

Maier, H. W. (2002). Role Playing: Structures and Educational objectives. *The International Child and Youth Care Network*, 36. Retrieved from http://www.cyc-net.org/cyc-online/cycol-0102-roleplay.html

Martínez-Riera, J. R., Sanjuán, A., Cibanal, L., & Pérez-Mora, M. J. (2011). Role-playing en el Proceso de Enseñanza-Aprendizaje de Enfermería: Valoración de los Profesores. *Cogitare Enfermagem*, 16(3), 411-417.

Mayer R. (1993). Diseño Educativo para un Aprendizaje Constructivista. *Journal of Educational Psychology*, 6, 153-169.

Mullan, B. A., & Kothe, E. J. (2010). Evaluating a Nursing Communication Skills Training Course: The Relationships between Self-rated Ability, Satisfaction, and Actual Performance. *Nurse Education in Practice*, 10(6), 374-378.

Naghavi, Z., Anbari, K., Saki, K., Mohtashami, A. Z., Lashkarara, G., & Derik, M. A. (2015). A Study on Effect of Training Communication Skills on Knowledge and Attitudes of Family Physicians and Patients Satisfaction. *Journal of Biology and Today's World*, 4 (1), 1-5

Nestel, D. & Tierney, T. (2007). Role-Play for Medical Students Learning about Communication: Guidelines for Maximising Benefits. *BMC Medical Educucation*, 7(3). Retrieved from http://www.biomedcentral.com/1472-6920/7/3.

Okuda, Y., Bryson, E. O., DeMaria, S., Jacobson, L., Quinones, J., Shen, B., & Levine, A. I. (2009). The Utility of Simulation in Medical Education: What is the Evidence? *Mount Sinai Journal of Medicine: A Journal of Translational and Personalized Medicine*, 76(4), 330-343.

Ramsden, P. (2003). *Learning to Teach in Higher Education*. London: Routledge.

Recker, M. M., Govindaraj, T., & Vasandani, V. (1998). Student Diagnostic Strategies in a Dynamic Simulation Environment. *Journal of Interactive Learning Research*, 9(2), 131–154.

San Martin, L., Utrilla, M., & Mediavilla, H. (2014). Efectividad de la Metodología Role-Playing como Herramienta de Enseñanza-Aprendizaje en el Grado de Odontología. In *Proceedins of XI Foro sobre la Educación de la Calidad de la Investigación y de la Educación Superior*. Bilbao: Deusto University.

Silverman, J., Kurtz, S. M., & Draper, J. (2013). *Skills for Communicating with Patients*. London: Radcliffe Pubublishing

Spitzberg, B. H. (2013). (Re)Introducing Communication Competence to the Health Professions. *Journal of Public Health Research*, 2(3), Retrieved from http://www.ncbi.nlm.nih.gov/pmc/articles/PMC4147740/

Stevenson K, Sander P. (2002). Medical Students are from Mars –Business and Psychology Students are from Venus –University Teachers are from Pluto? *Medical Teacher*, 24(1), 27-31.

Thorley, L., Gregory, R. D., & Gregory, R. (eds.). (1994). *Using Group-Based Learning in Higher Education*. London-Philadelphia: Psychology Press.

Van Ments M. (1989). *The Effective Use of Role Play*. New York: Nichols Publishing.

Yoshida, T., Milgrom, P., & Coldwell, S. (2002). How do US and Canadian Dental Schools Teach Interpersonal Communication Skills? *Journal of Dental Education*, 66(11), 1281-1288.